\documentclass[epj]{webofc}
\usepackage[varg]{txfonts}   
\usepackage{array} 
\usepackage{bm}

\wocname{\includegraphics[width=0.25cm,clip]{logo} ARENA-2016}
\woctitle{ARENA-2016}
\begin{document}
\title{Towards a cosmic-ray mass-composition study at Tunka Radio Extension}
%
%

\author{\firstname{D.}~\lastname{Kostunin}\inst{1}\fnsep\thanks{\email{dmitriy.kostunin@kit.edu}}
\and
\firstname{P.~A.}~\lastname{Bezyazeekov}\inst{2}
\and
\firstname{N.~M.}~\lastname{Budnev}\inst{2}
\and
\firstname{O.}~\lastname{Fedorov}\inst{2}
\and
\firstname{O.~A.}~\lastname{Gress}\inst{2}
\and
\firstname{A.}~\lastname{Haungs}\inst{1}
\and
\firstname{R.}~\lastname{Hiller}\inst{1}
\and
\firstname{T.}~\lastname{Huege}\inst{1}
\and
\firstname{Y.}~\lastname{Kazarina}\inst{2}
\and
\firstname{M.}~\lastname{Kleifges}\inst{3}
\and
\firstname{E.~E.}~\lastname{Korosteleva}\inst{4}
\and
\firstname{O.}~\lastname{Kr\"omer}\inst{3}
\and
\firstname{V.}~\lastname{Kungel}\inst{1}
\and
\firstname{L.~A.}~\lastname{Kuzmichev}\inst{4}
\and
\firstname{N.}~\lastname{Lubsandorzhiev}\inst{4}
\and
\firstname{R.~R.}~\lastname{Mirgazov}\inst{2}
\and
\firstname{R.}~\lastname{Monkhoev}\inst{2}
\and
\firstname{E.~A.}~\lastname{Osipova}\inst{4}
\and
\firstname{A.}~\lastname{Pakhorukov}\inst{2}
\and
\firstname{L.}~\lastname{Pankov}\inst{2}
\and
\firstname{V.~V.}~\lastname{Prosin}\inst{4}
\and
\firstname{G.~I.}~\lastname{Rubtsov}\inst{5}
\and
\firstname{F.~G.}~\lastname{Schr\"oder}\inst{1}
\and
\firstname{R.}~\lastname{Wischnewski}\inst{6}
\and
\firstname{A.}~\lastname{Zagorodnikov}\inst{2}
~(Tunka-Rex Collaboration) 
}

\institute{
Institut f\"ur Kernphysik, Karlsruhe Institute of Technology (KIT), Karlsruhe, Germany  
\and
Institute of Applied Physics, Irkutsk State University (ISU), Irkutsk, Russia  
\and
Institut f\"ur Prozessdatenverarbeitung und Elektronik, Karlsruhe Institute of Technology (KIT), Germany
\and
Skobeltsyn Institute of Nuclear Physics, Lomonossov University (MSU), Moscow, Russia
\and
Institute for Nuclear Research of the Russian Academy of Sciences, Moscow, Russia  
\and
Deutsches Elektronen-Synchrotron (DESY), Zeuthen, Germany
}

\abstract{%
The Tunka Radio Extension (Tunka-Rex) is a radio detector at the TAIGA facility located in Siberia nearby the southern tip of Lake Baikal. 
Tunka-Rex measures air-showers induced by high-energy cosmic rays, in particular, the lateral distribution of the radio pulses. 
The depth of the air-shower maximum, statistically depends on the mass of the primary particle, is determined from the slope of the lateral distribution function (LDF). 
Using a model-independent approach, we have studied possible features of the one-dimensional slope method and tried to find improvements for the reconstruction of primary mass.
To study the systematic uncertainties given by different primary particles, we have performed simulations using the CONEX and CoREAS software packages of the recently released CORSIKA~v7.5 including the modern high-energy hadronic models QGSJet-II.04 and EPOS-LHC. 
The simulations have shown that the largest systematic uncertainty in the energy deposit is due to the unknown primary particle.
Finally, we studied the relation between the polarization and the asymmetry of the LDF.
}
\maketitle
\section{Introduction}
In the last years the community made a large step forwards to understand the radio emission from extensive air-showers produced by high-energy (greater than 100~PeV) cosmic rays~\cite{Schroder:2016hrv}.
Using simulations made with the CoREAS software~\cite{Huege:2013vt}, several parameterizations describing lateral distributions of the radio amplitudes have been developed.
The contributions from the two main mechanisms of the generation of the signal (geomagnetic~\cite{KahnLerche1966}, and Askaryan or charge-excess~\cite{Askaryan1962a}) are either explicitly or implicitly included into all of these modern parameterizations.
Basically, they can be grouped by experiments applying them (in order of increasing number of free parameters): Tunka-Rex (4 parameters, asymmetry $\varepsilon$ is explicitly included into parameterization)~\cite{Kostunin:2015taa}; LOFAR, and later AERA (7 parameters, describing two gaussian functions shifted respective to each other)~\cite{Nelles:2014xaa}; and the straight-forward LOFAR approach (full simulation of radio footprints for each antenna position)~\cite{Buitink:2014eqa}.

The asymmetry caused by interference between the Askaryan effect and the geomagnetic effect has already been treated within all modern approaches, and has been measured by means of the polarization study~\cite{Aab:2014esa,Schellart:2014oaa}.
In the present paper we will discuss its impact on the lateral distribution and the connection of the asymmetry with the shower maximum.

First, we study the model-independent behavior of the radio emission, then we check the influence and uncertainties given by different hadronic models.
Finally, we discuss possible improvements for the reconstruction of the shower maximum.
\section{Model-independent considerations and model-dependent uncertainties}



Let us study the LDF using the following assumptions: the distribution of the electrons behaves as Gaisser-Hillas function and the density of the Earth's atmosphere falls exponentially with increasing altitude, namely we use the CORSIKA parameterization of the standard atmosphere~\cite{HeckKnappCapdevielle1998}.
The simple form of the amplitude of the radio signal ${\mathcal{E}_\nu(r>r_c)}$ beyond the Cherenkov bump $r_c$ with frequency $\nu$ at distance $r$ from the shower axis is~\cite{Allan1971}:
\begin{equation}
\label{eq1}
\mathcal{E}_\nu(r>r_c) = \kappa \int\limits^{h^\nu_2(r,n_\mathrm{r})}_{h^\nu_1(r,n_\mathrm{r})} \frac{N(h)}{h} \mathrm{d}h \propto \exp \left(-\frac{(r/r_1)^{\alpha_1}}{h_\mathrm{max}}f_{\mathrm{int}}(h_\mathrm{max},...)\right)\,,\,\,\, h^\nu_{1,2}(r,n_\mathrm{r}) = \left(\frac{r}{r_{1,2}(\nu,n_\mathrm{r})}\right)^{\alpha_{1,2}(\nu,n_\mathrm{r})}\,,
\end{equation}
where $\kappa$ is the normalization coefficient (the dependence on geomagnetic angle $\alpha_\mathrm{g}$ has already been taken into account),
$N(h)$ is the number of electrons at the altitude $h$, and $h^\nu_{1,2}(r,n_\mathrm{r})$ are the integration limits depending on the distance to shower axis $r$ and refractive index $n_\mathrm{r}$.
In other words, this integral describes what parts of the air-shower contribute to the signal $\mathcal{E}_\nu(r)$.
The curves denoting the behavior of $h^\nu_{1,2}(r,n_\mathrm{r})$ are presented in Ref.~\cite{KostuninISVHECRI2016}. 
One can see from Eq.~(\ref{eq1}), that the height of the shower maximum $h_\mathrm{max}$ is encoded in the slope of the LDF. 
Simplifying ${f_\mathrm{int} = 1}$ one still conserves the correlation between $\mathcal{E}$ and $h_\mathrm{max}$ (see Ref.~\cite{KostuninISVHECRI2016}), 
what means, that the radio signal has a high sensitivity to the position of the shower maximum.
On the other side, this one-dimensional slope method does not give additional information on the type of primary particle.

To study uncertainties given by hadronic interactions and shower-to-shower fluctuations, we performed simulations with the recently released CORSIKA~v7.5.
Both, QGSJET-II.04 and EPOS-LHC yield almost the same radio amplitude with the difference less than one given by shower-to-shower fluctuations.
They both show a shift of 12\% between proton and iron in the radio energy deposit, which is much larger than the effect of shower-to-shower fluctuations for these particles: 5\% and 1.3\% for proton and iron, respectively.
This uncertainty can be reduced by taking into account information from muon detectors, in our case we plan to combine measurements of Tunka-Rex and Tunka-Grande~\cite{TunkaRex_NIM_2015,SchroederArena2016}. 





\section{Hints from the charge-excess asymmetry}

\begin{figure}[t!]
\begin{center}
\includegraphics[width=1.0\linewidth]{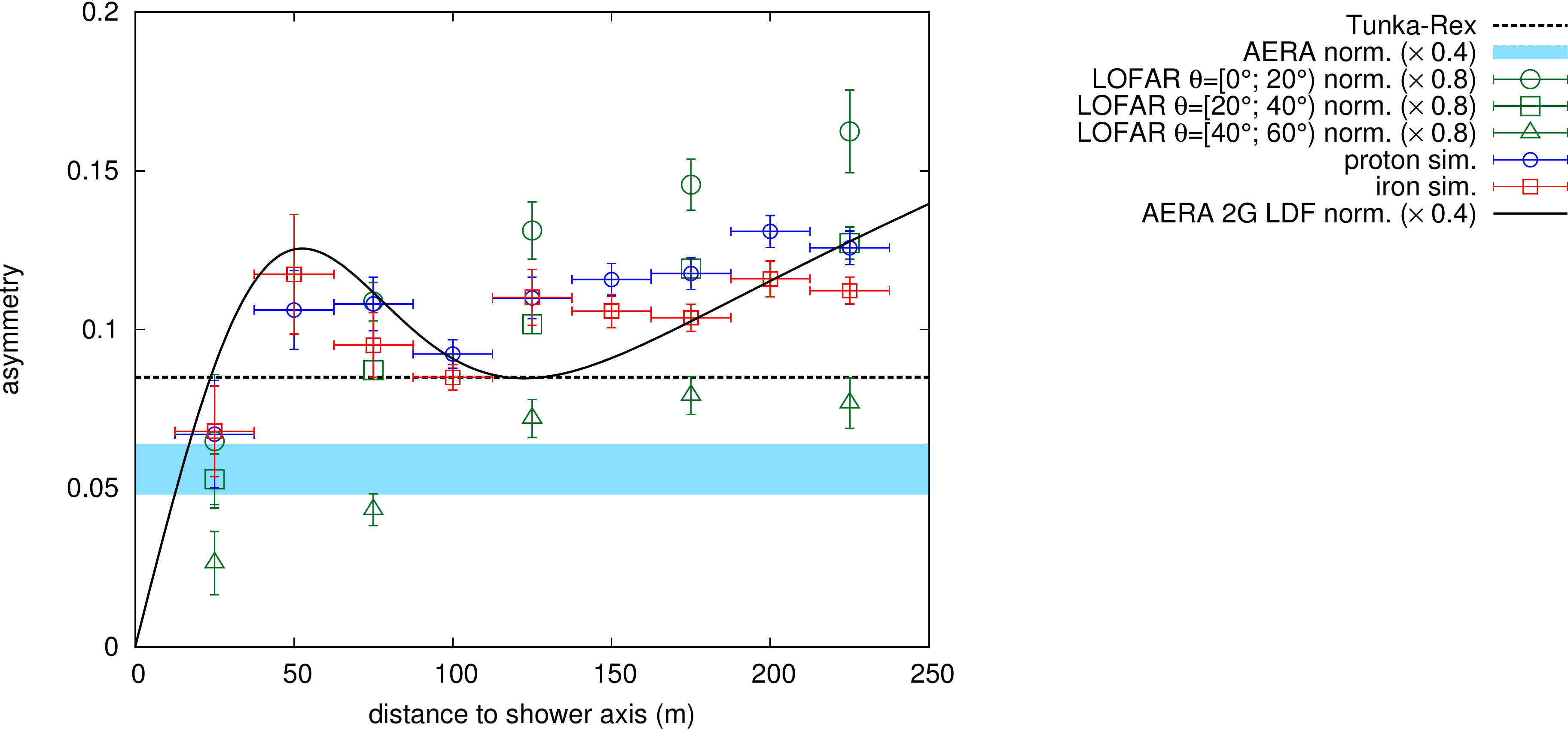}
\caption{Askaryan asymmetry $\varepsilon(r)$ (Eq.~\ref{eq_eps}) normalized to a geomagnetic field in Tunka Valley as a function of distance to the shower axis. 
Points indicate polarization measurements be LOFAR~\cite{Schellart:2014oaa} (green) and CoREAS simulations~\cite{Kostunin:2015taa} (blue and red).
Black solid line indicates LDF asymmetry $\varepsilon(r)$ from Eq.~(\ref{eq_eps}).
The blue band is the polarization measurements by AERA~\cite{Aab:2014esa} (with uncertainties), the dashed line is the theoretical prediction for Tunka-Rex~\cite{Kostunin:2015taa}.
} 
\label{asymm_profile}
\end{center}
\end{figure}

In the work~\cite{Kostunin:2015taa} it is shown that the charge-excess asymmetry has a non-trivial dependence on the distance to the shower axis, 
particularly, the asymmetry features a local maximum depending on the distance to the shower maximum (see Fig.~\ref{asymm_profile}).
This structure was obtained by analyzing the polarization of CoREAS simulations at individual antenna positions.
In the present section we discuss the structure of the asymmetry by means of a completely different approach: we try to extract it by combining the Tunka-Rex~\cite{Kostunin:2015taa} and AERA~\cite{Aab:2015vta} LDF parameterizations of the amplitude $\mathcal{E}$ depending on the coordinate $\bm{r}$ in the shower plane with ${\hat e}_x$ the direction of the geomagnetic Lorentz force.

Let us write down the AERA parameterization:
\begin{equation}
\mathcal{E}_{\mathrm{2G}}(\bm{r}) = A \Biggr[ \, \exp\left(\frac{-(\bm{r} + C_1 \, \bm{\mathrm{\hat e}}_x)^2}{\sigma^2}\right) - C_0 \, \exp\left(\frac{-(\bm{r} + C_2 \, \bm{\mathrm{\hat e}}_x)^2}{(C_3 e^{C_4 \, \sigma})^2}\right) \Biggr]\,,
\label{aera_ldf}
\end{equation}
parameters of which are described in Ref.~\cite{Aab:2015vta},
and the Tunka-Rex parameterization: 
\begin{equation}
\mathcal{E}_{\phi}(r,\phi_g) = \mathcal{E}_0(r) \sqrt{\sin^2\alpha_g + \varepsilon^2(r) + 2\varepsilon(r)\cos\phi_g}\,,
\end{equation}
with parameters described in Ref.~\cite{Kostunin:2015taa}.
One can see, that the asymmetry term $\varepsilon(r)$ is in Eq.~(\ref{aera_ldf}) only implicit.
The asymmetry appears along the Lorentz force, thus we define $\bm{r}_x = (r,0)$:
\begin{equation}
\varepsilon(r) = \sin\alpha_g\frac{{\mathcal{E}_{\mathrm{2G}}(r)} - \tilde{\mathcal{E}}_{\mathrm{2G}}(r)}{\mathcal{E}_{\mathrm{2G}}(r) + \tilde{\mathcal{E}}_{\mathrm{2G}}(r)}\,,\,\,\,
\begin{cases}
\mathcal{E}_{\mathrm{2G}}(r) = \mathcal{E}_{\mathrm{2G}}(\bm{r}_x) \\
\tilde{\mathcal{E}}_{\mathrm{2G}}(r) = \mathcal{E}_{\mathrm{2G}}(-\bm{r}_x) \\
\end{cases}.
\label{eq_eps}
\end{equation}
We take $C_0$ -- $C_4$ from Ref.~\cite{Aab:2015vta} for a zenith angle of $\theta = 45^{\circ}$, a typical geomagnetic angle for this zenith of ${\alpha_\mathrm{g} = 45^\circ}$ and a width of the main gaussian ${\sigma = (-2a_2(E, \theta))^{-1/2} = 152.32}$~m, where ${E = 1}$~EeV, and $a_2(E,\theta)$ is defined in Ref.~\cite{Bezyazeekov:2015ica}.
By this calculation, the Askaryan asymmetry of the polarization can be directly compared with the Askaryan asymmetry of the lateral distribution.
The comparison is presented in Fig.~\ref{asymm_profile}.
One can see, that both definitions are in good agreement, which leads to an interesting conclusion: the asymmetry (or charge-excess) information can be extracted from the more simple measurement of the total radio amplitude, instead of precise measurements of the components of the electrical field.
Measuring the amplitude asymmetry requires higher number of stations per events, but lower signal-to-noise ratios.
Moreover, as it was shown in Ref.~\cite{Kostunin:2015taa}, the behavior of the asymmetry is connected to the distance to shower maximum,
i.e. an accurate measurement of the asymmetry by either means should be sensitive to the mass composition.
Finally, the asymmetry contains information not only on the total number of the charge particles, but also on the dynamics of their creation.

\section{Summary}
In the present work we have studied ideas how to better reconstruct the primary mass using radio measurements.
Since radio detectors feature calorimetric measurements, the main property related to mass composition is the shower maximum.
We have shown with simple calculations, that the Gaisser-Hillas distribution of the particle cascade produces an exponential-like lateral distribution of the amplitudes far from the shower axis, and the position of the shower maximum is directly encoded in the slope of this exponent.
However, radio is sensitive only to the electromagnetic component of air-showers, which leads to an additional uncertainty of 6\% in energy estimation due to unknown primary particle.

As an alternative to the slope method we have studied other properties of the LDF.
We found that the asymmetry, which is slightly sensitive to shower maximum, can be equally well determined from amplitude and polarization measurements.
We now plan to introduce this study to our standard analysis of the recently upgraded Tunka-Rex array.

%

\section*{Acknowledgements}
The construction of Tunka-Rex was funded by the German Helmholtz association and the Russian Foundation for Basic Research (grant HRJRG-303). 
Moreover, this work has been supported by the Helmholtz Alliance for Astroparticle Physics (HAP), by Deutsche Forschungsgemeinschaft (DFG) grant SCHR 1480/1-1, and by the Russian grant RSF 15-12-20022. 

\bibliography{references}

\end{document}